\documentstyle[aps,prl,epsfig,multicol]{revtex}
\input psfig.sty
\newcommand{\etal}{{\it et al.}}
\newcommand{\np}{Nucl. Phys.}
\newcommand{\prev}{Phys. Rev.}

\begin{document}




\title{Measurement of the Neutron Spin Structure Function ${\bf g_1^n}$ \\
  with a Polarized $^3$He Internal Target\\}
\author{
\centerline {\it The HERMES Collaboration}\medskip \\
K.~Ackerstaff$^{5}$,
A.~Airapetian$^{30}$,
I.~Akushevich$^{6}$,
N.~Akopov$^{30}$, 
M.~Amarian$^{30}$, 
E.C.~Aschenauer$^{12,22}$,
R.~Avakian$^{30}$,
H.~Avakian$^{10}$, 
A.~Avetissian$^{30}$, 
B.~Bains$^{14}$
% ...
\medskip\\}
\address{
% ...
$^{5}$DESY, Deutsches Elektronen Synchrotron, 22603 Hamburg, Germany\\
$^{6}$DESY-IfH Zeuthen, 15738 Zeuthen, Germany\\
% ...
$^{10}$Istituto Nazionale di Fisica Nucleare, Laboratori Nazionali di Frascati, 00044 Frascati, Italy\\
% ...
$^{12}$University of Gent, 9000 Gent, Belgium\\
% ...
$^{14}$University of Illinois, Urbana, Illinois 61801, USA\\
% ...
$^{22}$Nationaal Instituut voor Kernfysica en Hoge-Energiefysica (NIKHEF), 1009 DB Amsterdam, The Netherlands\\
% ...
$^{30}$Yerevan Physics Institute, 375036, Yerevan, Armenia\\
}
\date{March 3, 1997}
\maketitle

\begin{abstract}
Results are reported from the HERMES experiment at HERA on a 
measurement of the neutron spin structure function 
$g_1^n(x,Q^2)$  in deep inelastic scattering 
using 27.5~GeV longitudinally polarized positrons incident on 
a polarized $^3$He internal gas target. The data cover the kinematic range 
$0.023<x<0.6$ and 1~(GeV/c)$^2<Q^2<15$~(GeV/c)$^2$. 
The integral $\int_{0.023}^{0.6} g_1^n(x)\,
dx$ evaluated at a fixed $Q^2$ of 2.5~(GeV/c)$^2$ is $-0.034\pm
0.013$(stat.)$\pm 0.005$(syst.). Assuming Regge behavior at low $x$,
the first moment $\Gamma_1^n=\int_0^1 g_1^n(x)\, dx$ is 
$-0.037\pm 0.013$(stat.)$\pm 0.005$(syst.)$\pm 0.006$(extrapol.).

\end{abstract}

\begin{multicols}{2}[]
Deep inelastic scattering with lepton beams is an 
important tool for understanding the quark-gluon structure of the nucleon.
With polarized beams and targets, the spin structure of the nucleon 
is probed via scattering of charged leptons from nucleons through the
exchange of virtual photons (with energy transfer $\nu$ and squared
four momentum $-Q^2$). Inclusive polarized 
scattering is characterized by two structure functions:
$g_1(x,Q^2)$ and $g_2(x,Q^2)$, with $x = Q^2/2M\nu$. 
The structure function $g_1$ is determined from the virtual photon asymmetries
$A_1$ and $A_2$ via
\begin{equation}
g_1 = {F_1 \over (1 + \gamma^2)}[A_1 + \gamma A_2],
\label{g1eq}
\end{equation}
where $F_1$ is the unpolarized structure function and $\gamma^2 = Q^2/\nu^2$.

In the quark parton model the structure function $g_1$ is related to the quark
spin distributions $\Delta q_f$ through
\begin{equation}
g_1(x,Q^2) = {1 \over 2}\sum_f e_f^2 \Delta q_f(x,Q^2),
\label{g1spin}
\end{equation}
where the sum is over quark flavors; $e_f$ is the quark charge in units of
the elementary charge $e$, and $x$ is interpreted as 
the fraction of the nucleon's light 
cone momentum carried by the struck quark.
Important information can be obtained from the structure function integrals
for proton and neutron
$\Gamma_1^p = \int_0^1 g_1^p(x)dx$ and $\Gamma_1^n = \int_0^1 g_1^n(x)dx$.
The fundamental Bjorken sum rule relates the difference 
$\Gamma_1^p - \Gamma_1^n$ to the weak axial charge $g_A$ \cite{bjsr}.
Including QCD corrections~\cite{qcd4a,qcd4}, this sum rule has been 
confirmed experimentally within about 10 percent,
while the model dependent Ellis-Jaffe sum rules \cite{ejsr}, which provide 
predictions for $\Gamma_1^p$ and $\Gamma_1^n$ separately, are significantly
violated \cite{emc1,e142n1,e142n2,smc1,smc4,e143d,smc2,smc3,e143p,smcpre1,smcpre2}. 
The experimental results suggest that only a fraction
of the nucleon's spin is due to the quark spins, and that 
the remainder results from gluons or orbital angular momentum. 

HERMES is based on two novel techniques: an internal gas
target of polarized atomic hydrogen, deuterium or $^3$He, and a 
high current longitudinally polarized positron beam circulating in
a high-energy storage ring.
A measurement of $g_1^n$ extracted from data taken during the initial 
running of the HERMES experiment is reported in this letter. 
The experiment utilized
a 27.5 GeV beam of longitudinally polarized positrons in the HERA storage
ring at DESY incident on a longitudinally polarized $^3$He internal 
target \cite{hermes1,hermes2,specpaper}. 

The positron beam in the HERA ring becomes transversely
polarized to a high level by the Sokolov-Ternov
mechanism~\cite{sok64}. The time constant for this process depends
on the beam energy and the beam tune and is approximately 20-25
minutes. Precise alignment of the
machine quadrupoles and fine tuning of the orbit parameters are needed
to achieve high polarisation. The required longitudinal polarization 
direction is obtained using spin rotators located upstream and downstream
of the HERMES experiment in the East straight section of HERA.
This results in the first longitudinally polarized electron beam in a 
high-energy storage ring ~\cite{bar95}. The transverse polarization 
was measured continuously using Compton backscattering of circularly 
polarized laser light.
Values of the equilibrium polarization in the range 40\% to 65\%
were obtained under normal running conditions.  Experimental data were analyzed
only when the polarization was above 40\% to reduce sensitivity to 
systematic effects. 
The average polarization for the analyzed data was 55\%. 
The fractional statistical error for a single 
60 s polarization measurement was typically 1-2\% and the overall
fractional systematic error was 5.4\%, dominated by the uncertainty in the 
calibration of the beam polarimeter.
A single beam polarization direction was used for the present measurements.

The $^3$He target atoms were polarized by spin exchange collisions with
$^3$S metastable $^3$He atoms in a glass cell.
The $^3$S atoms were produced by a weak RF
discharge and polarized by optical pumping with 1083 nm 
laser light~\cite{ref1}.  
The polarized atoms diffused 
from the glass pumping cell into a 400 mm long open ended thin-walled storage
cell inside the positron ring~\cite{ref1prime}.
The storage cell was constructed of 125 $\mu$m
thick ultrapure aluminum and was cooled to typically 
25~K~\cite{ref2}. This provided a target of pure
atomic species with an areal density of approximately
$3.3 \times 10^{14}$ atoms/cm$^2$.
The polarization direction was defined by a 3.5 mT magnetic 
field parallel to the beam direction. 

The polarization of
the $^3$He gas in both the pumping and the storage cells 
was measured continuously with optical polarimeters~\cite{ref3}.
The nuclear polarization in the
pumping cell was determined from the polarization of emitted photons
produced via atomic excitation by the RF discharge. A second polarimeter
monitored the polarization in the storage cell by
measuring the polarization of photons emitted from atoms that were excited
by the positron beam. These
measurements were used to investigate the possibility that atoms
were depolarized in the storage cell. No evidence for such effects
was found. Cell wall depolarization was measured at lower temperatures
using this technique in dedicated test measurements ~\cite{ref4}.  The average
value of the target polarization during the experiment was 46\% with
a fractional uncertainty of 5\%. 
The target polarisation direction was reversed every 10 min by
reversing the laser helicity.

The luminosity was measured by detecting Bhabha scattered
target electrons in coincidence with the scattered positrons in a pair of 
NaBi(WO$_4$)$_2$ electromagnetic calorimeters. 
During the course of one positron fill (typically 8 hr), the positron
current
in the ring decreased from typically 30 mA at injection to $\sim 10$ mA, at 
which point the positron beam was dumped. 

\begin{figure}
\epsfxsize 16 cm \epsfbox{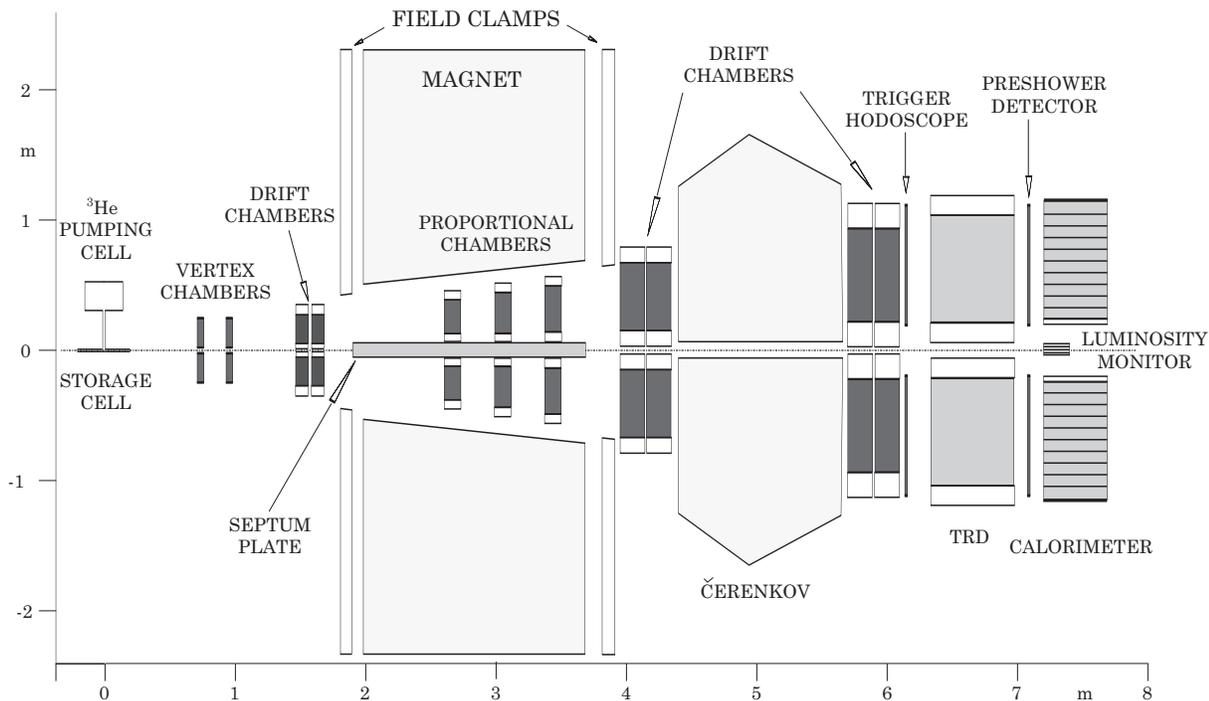}
\caption{Schematic diagram of the experimental apparatus (side view).
\label{detplot}}
\end{figure}

A schematic diagram of the apparatus~\cite{specpaper} is shown in
Fig.~\ref{detplot}.  It consists of a large dipole magnet 
surrounding the positron and proton beam pipes of HERA. 
The beam is shielded from the spectrometer's magnetic field 
by a horizontal iron plate. 
The spectrometer is constructed as two identical halves, mounted above
and below the region of the beam pipes
and the horizontal iron plate.
The scattering angle acceptance of the spectrometer
is 40 mrad $< \theta <$ 220 mrad. 
Each half contains thirty-six drift chamber planes for tracking. 
A pattern-matching algorithm and momentum look-up method \cite{wwc} provides 
fast track reconstruction.
A momentum resolution of 1 - 2\% for positrons, depending on kinematics,
and an average angular resolution of 0.6 mrad was achieved. 
The trigger was formed from a coincidence between a pair of scintillator
hodoscope planes and a Pb-glass calorimeter.  
The trigger required an energy of greater than $3.5$ GeV deposited in the  
calorimeter, resulting in a typical event rate of 50 Hz. 
Positron identification was accomplished using the calorimeter, the second
hodoscope, which was preceded by 2 radiation lengths of Pb and 
functioned as a preshower counter, 
six transition radiation detector modules,
and a N$_2$ threshold gas Cerenkov counter. This system provided positron
identification with an average of 98\% efficiency and a hadron 
contamination $< 1\%$. 
A two-stage collimator system
mounted upstream of the target cell provided shielding from the
synchrotron radiation generated in the beam bending and focussing
components and from beam-halo positrons. 
With proper beam tuning, the detector system was
essentially free of electromagnetic background from such processes. 
A negligible number of triggers from particles scattering
from the storage cell walls were observed.

The structure function $g_1^n$ was extracted from the measured longitudinal 
asymmetry $A_{||}$ of the scattering cross section
using Eq.~(\ref{g1eq}) with $A_1 = A_{||}/D - \eta A_2$ and 
$F_1 = F_2 (1+\gamma^2)/(2x(1+R))$. Here
$D = [1-(1-y)\epsilon]/(1+\epsilon R)$ is the virtual photon 
depolarization factor, 
$\epsilon = [4(1-y) - \gamma^2 y^2]/[2y^2+4(1-y)+\gamma^2 y^2]$ is the 
degree of transverse polarization of the virtual photon,
$\eta = \epsilon \gamma y/[1-\epsilon(1-y)]$ , 
$R = \sigma_L/\sigma_T$ is the longitudinal-to-transverse
virtual photon cross-section ratio, 
$y = \nu/E$ and $E$ is the beam energy. 
The magnitude of $A_2$ is constrained to be less than  $\sqrt{R}$ 
and has been 
measured previously \cite{e142n2} to be consistent with zero within
large errors. Thus its 
contribution to $g_1$ was neglected but its uncertainty was included 
as a systematic error.

The value of $A_{||}/D$ for $^3$He was extracted from the measured 
counting rates using the formula 
\begin{equation}
{A_{||} \over D} = {N^-L^+ - N^+L^- \over D(N^-L^+_P + N^+L^-_P)},
\label{eq:ap}
\end{equation}
where $N^+ (N^-)$ is the counting rate for target spin parallel
(anti-parallel) 
to the beam spin, $L^{\pm}$ are the deadtime-corrected luminosities
for each target spin state, and $L^{\pm}_P$ are the deadtime-corrected
luminosities weighted by the product of the magnitudes
of the beam and target polarizations for each spin state. This quantity was
binned in $x$ and $y$ to take into account the strong variation of $D$
with $y$ and was determined separately for each positron fill.

After applying data quality criteria and kinematic cuts 
($Q^2 > 1$ (GeV/c)$^2$, $W^2 > 4$ (GeV/c)$^2$ and $y < 0.85$)
$2.7 \times 10^6$ events
were available for the asymmetry analysis. 
Corrections were applied to account for background from charge symmetric
processes (eg. $\gamma \rightarrow e^+ e^-$), and from misidentified
hadrons. Whereas the asymmetry of the former source was consistent 
with zero, the latter exhibited a non-zero asymmetry which was typically
20-50\% of the positron asymmetry.
The corrections from both backgrounds were at most 2\% of the asymmetry 
for the smallest
$x$ bins and were negligible for large $x$ values.
QED radiative corrections were applied 
using the standard procedure \cite{radcor},
with corrections of typically 20\% of the observed asymmetry. 
Monte Carlo simulations showed that smearing corrections 
due to the finite resolution of the spectrometer are negligible. 

Corrections for nuclear effects are required
to determine the neutron structure function $g_1^n$ 
with a $^3$He target. The
wave function for $^3$He is dominated by the configuration with the
protons paired to zero spin. Therefore, 
most of the asymmetry from $^3$He is due to the neutron \cite{friar}.
A detailed calculation~\cite{nuclcorr} shows that binding effects
and Fermi motion are negligible in extracting $g_1^n$ from polarized 
$^3$He data, and that nuclear effects can be largely treated as a 
dilution due to scattering from the two largely unpolarized protons. 
A correction for the non-zero polarization of the protons 
($-0.028 \pm 0.004$), using the E-143 results for $A_1^p$ \cite{e143p},
and the neutron polarization ($0.86 \pm 0.02$) \cite{nuclcorr} 
has been included. 

\begin{figure}[ht]
\begin{center}
  \epsfxsize 15 cm \epsfbox{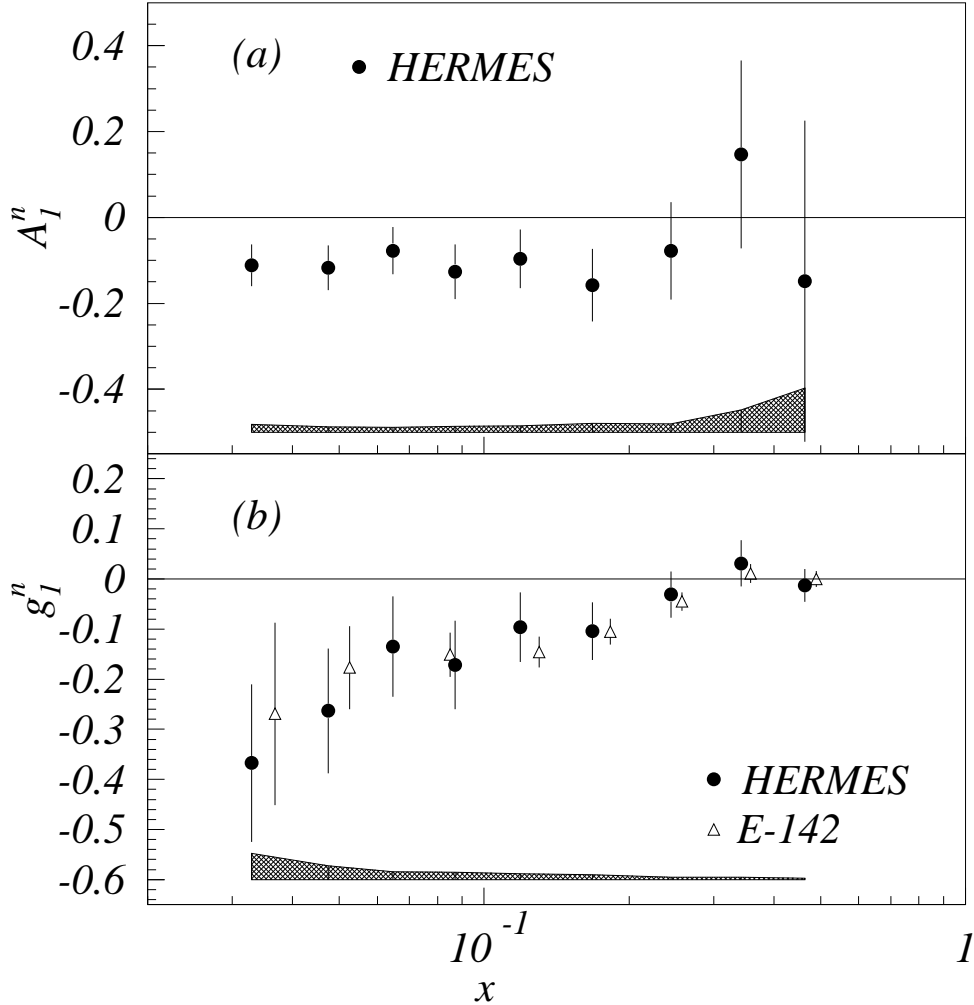}
\caption{The spin asymmetry $A_1^n$ (a) and the spin structure
  function $g_1^n$ (b) of the neutron as a function
  of $x$. The values are given
  for the measured $\left<Q^2\right>$. 
  The error bars are statistical uncertainties. The error bands
  show the systematic uncertainties. The data points from E-142 have
  been displaced slightly in $x$ for comparison with the present 
  experiment.
\label{g1plot}}
\end{center}
\end{figure}

The extracted virtual photon asymmetry $A_1^n(x)$ is presented in
Fig.~\ref{g1plot}(a).
The averaged kinematic quantities and asymmetry results are listed
in Table~\ref{g1table}.
The corresponding $g_1^n(x)$, extracted from equation (\ref{g1eq}) and
using parameterizations of the unpolarized structure function
$F_2$ \cite{f2ref} and $R$ \cite{rref}, is shown in Fig.~\ref{g1plot}(b) and
compared with the previous $^3$He experiment from SLAC E-142~\cite{e142n2}. 
The systematic uncertainties in the present experiment
are small compared to the statistical
uncertainties, as indicated by the error bands in Fig.~\ref{g1plot}.
The statistical uncertainties have been extracted from the observed
fluctuations of the positron yields and exceed the uncertainty calculated
from the number of events by 10\%.
The dominant sources of systematic errors on $A_{||}/D$
are the uncertainties in the measured beam and target polarization.
In addition for $g_1^n$ 
there are contributions of similar size which result from 
uncertainties in radiative corrections and nuclear corrections,
and smaller contributions from the uncertainties in the knowledge of
$F_2^n$, $A_2$ and $R$.

For an evaluation of the Ellis-Jaffe sum rule the integral of 
$g_1^n(x)$ must be determined at a fixed $Q^2$
and an extrapolation into the unmeasured $x$ regions must be made. 
This requires an assumption
on the $Q^2$ dependence of either $A_1$ or $g_1$.
To evolve $g_1$ to a fixed value of $Q_0^2 = 2.5$ (GeV/c)$^2$, which is 
near to the mean $Q^2$ value of the data ($2.3 (GeV/c)^2$), the assumption
that $A_1$ is independent of $Q^2$ over the limited $Q^2$ range of our 
data has been used. This assumption is consistent with existing data
\cite{e143q2}. Next-to-Leading-Order (NLO) QCD evolution \cite{qcd4,qcdevo2}
gives a slightly different result that changes the integral of $g_1^n$ 
over the measured $x$ range by $< 5\%$.
Including this difference in the systematic error yields
$\int_{0.023}^{0.6} g_1^n(x,Q_0^2)\, dx =
-0.034\pm 0.013$(stat.)$\pm 0.005$(syst.). 

For the large $x$ extrapolation, we used a parametrization for 
$F_2^n$~\cite{f2ref} and assumed several models for
the behavior of $A_1^n$ for $x > 0.6$. Since
$A_1^n$ is expected \cite{bigx} to approach unity for $x \rightarrow 1$,
we considered a linear increase for $A_1^n$ from 0 at $x = 0.6$ to 
1 at $x = 1$ as well as the parameterization of ref.~\cite{schaef}. 
These studies indicate that 
$\int_{0.6}^1 g_1^n(x)dx = 0.002 \pm 0.003$.
For the low $x$ extrapolation there is no clear prediction. 
For comparison with previous measurements
\cite{e142n2,smc1,smc4,smc2,smc3}, 
we quote the integral $\Gamma_1^n$ assuming 
a simple Regge parameterization at low $x$ \cite{reggeexp,regge} of
$g_1 \propto x^{-\alpha}$ with $\alpha$ in the range $-0.5 - 0$ fitted
to the data for $x < 0.1$. This gives
$\int_0^{0.023} g_1^n(x)dx = -0.005 \pm 0.005$, where 
a 100\% uncertainty has been assigned to the value.
However, it should be noted that recent work \cite{qcd4} indicates
that a NLO treatment of the low $x$ region could 
yield different results for the low $x$ extrapolation.
Combining the contributions 
from different $x$ regions leads to a total integral
of $\int_0^1 g_1^n(x,Q_0^2)dx = 
-0.037\pm 0.013$(stat.)$\pm 0.005$(sys.)$\pm 0.006$(extrapol.) 
in good agreement with the value from experiment E-142~\cite{e142n2} using 
$^3$He and the SMC~\cite{smcpre1,smcpre2} and E-143~\cite{e143p,e143d}
experiments using the difference of deuteron and proton.

In summary, 
the neutron spin structure function $g_1^n$ has been measured with a 
polarized $^3$He target. 
The results are in agreement with those of the SLAC E-142 
experiment, but have been determined with an entirely new technique --
a windowless polarized internal target with 
pure atomic species in a positron storage ring.
Semi-inclusive asymmetries extracted from the present 
data set will be presented in a future publication. 

We gratefully acknowledge the DESY management for its support and 
the DESY staff and the staffs of the
collaborating institutions for their significant effort. This work was
supported by the U.~S.~Department of Energy and National Science
Foundation; 
the Natural Sciences and Engineering Research Council of Canada; 
the German Bundesministerium f\"ur Bildung, Wissenschaft Forschung 
und Technologie, the German Academic Exchange Sevice (DAAD); 
the Italian Istituto Nazionale di Fisica Nucleare; 
Monbusho International Scientific Research Program, JSPJ, and Toray
Science Foundation of Japan;
the Dutch Foundation for Fundamental Research of Matter;
the NFWO/IIKW, Belgium and 
the INTAS contribution from the European Community, 




\end{multicols}
\clearpage

\begin{table}
\caption{\small Results on $A_1^n(x)$ and $g_1^n(x)$ at the measured $Q^2$.
\label{g1table}}\medskip
\begin{tabular}{ccccc}
x-range & $\left<x\right>$ & $\left<Q^2\right>$ [(GeV/c)$^2$] & $A_1^n\pm$stat.$\pm$syst. &$g_1^n\pm$stat.$\pm$syst. \\
\hline
0.023-0.040 &0.033  &1.22  & $-0.111\pm 0.048\pm 0.018$ & $-0.367\pm 0.157\pm 0.052$ \\
0.040-0.055 &0.047  &1.47  & $-0.117\pm 0.052\pm 0.013$ & $-0.263\pm 0.124\pm 0.028$ \\
0.055-0.075 &0.065  &1.73  & $-0.077\pm 0.055\pm 0.011$ & $-0.135\pm 0.100\pm 0.016$ \\
0.075-0.10~ &0.087  &1.99  & $-0.126\pm 0.064\pm 0.014$ & $-0.172\pm 0.088\pm 0.015$ \\
0.10-0.14   &0.119  &2.30  & $-0.097\pm 0.068\pm 0.015$ & $-0.096\pm 0.069\pm 0.012$ \\
0.14-0.20   &0.168  &2.65  & $-0.158\pm 0.085\pm 0.020$ & $-0.104\pm 0.057\pm 0.010$ \\
0.20-0.30   &0.244  &3.07  & $-0.078\pm 0.113\pm 0.019$ & $-0.031\pm 0.046\pm 0.005$ \\
0.30-0.40   &0.342  &3.86  & $+0.146\pm 0.219\pm 0.052$ & $+0.031\pm 0.046\pm 0.005$ \\
0.40-0.60   &0.464  &5.25  & $-0.149\pm 0.374\pm 0.103$ & $-0.013\pm 0.033\pm 0.003$ 

\end{tabular}
\end{table}

\end{document}